\documentstyle[12pt]{article}
\topmargin - 10 mm
\begin{document}

\title{Buckling of rods with spontaneous twist and curvature}

\author{A.D. Drozdov$^{1}$ and Y. Rabin$^{2}$\\
$^{1}$ Institute for Industrial Mathematics\\
4 Hanachtom Street, Beersheba, 84311 Israel\\
$^{2}$ Department of Physics, Bar-Ilan University\\
Ramat-Gan, 52900 Israel}

\date{}
\maketitle

\begin{abstract}
We analyze stability of a thin inextensible elastic rod
which has non-vanishing spontaneous generalized torsions in its
stress-free state.
Two classical problems are studied, both involving spontaneously twisted
rods: a rectilinear rod compressed by axial forces,
and a planar circular ring subjected to uniform radial pressure
on its outer perimeter.
It is demonstrated that while spontaneous twist stabilizes a rectilinear
rod against buckling, its presence has an opposite effect on a closed ring.
\end{abstract}

The history of the buckling instability in a thin elastic rod under
compression goes back to the works of Euler and his contemporaries
\cite{Lov44},
who laid down the foundations of the general theory of elastic stability
\cite{TG63}.
Although this field has long been the domain of engineers and applied
mathematicians,
there has recently been a renaissance of interest in problems related to
the stability of thin filaments in the theoretical physics community
\cite{GL92}--\cite{GF00} prompted by experimental advances in the art
of mechanical manipulation of single DNA molecules and rodlike protein
assemblies \cite{SCB96}--\cite{FESL97}.
Because double stranded DNA molecules are helices that contain strongly
curved regions \cite{BFKS95}, their modeling as elastic rods has to take
into account their spontaneous curvature and twist.
While general considerations concerning such ``naturally'' curved rods
can already be found in the work of Kirchhoff \cite{Lov44},
little is known about the effect of spontaneous curvature and
twist on the stability with respect to buckling.
The present study deals with the effect of spontaneous twist
on the static stability of a thin inextensible elastic rod.
Nonlinear equilibrium equations are derived for a rod under the action
of external forces and moments.
Two classical problems are analyzed which have well-known solutions in
the absence of spontaneous twist \cite{Lov44}:
a rectilinear rod compressed by forces applied to its ends
and a ring compressed by radially inwards directed pressure
uniformly distributed along its perimeter.
It is demonstrated that while spontaneous twist stabilizes the straight
rod against buckling, it has a destabilizing effect on the ring.
In both cases buckling takes place through a three-dimensional instability.

\section{Equilibrium equations}
Consider an elastic rod of length $L$ and cross-section ${\cal S}$ which
has two axes of symmetry.
We assume that $L$ $\gg a$ where $a$ is the largest dimension associated
with the cross-section.
Denote by $s$ the arc-length which parameterizes the centerline of the
rod in the stress-free reference configuration.
The radius vector ${\bf R}_{0}(s)$ describes the spatial
position of the centerline and ${\bf t}_{0}=d{\bf R}_{0}/ds$ the unit
tangent vector in the reference state.
The unit vectors ${\bf u}_{0}(s)$ and ${\bf v}_{0}(s)$ are directed along
the axes of symmetry of the cross-section.
We would like to emphasize that ${\bf u}_{0}$ and ${\bf v}_{0}$
are associated with physical attributes of rods of small but non-vanishing
thickness and, in general, do not coincide with the normal and the bi-normal
vectors which appear in the description of three-dimensional
configurations of infinitely thin geometrical lines.
Differentiation of the vectors ${\bf t}_{0}$, ${\bf u}_{0}$
and ${\bf v}_{0}$ with respect to $s$ results in
\begin{equation}
\frac{d{\bf t}_{0}}{ds}=\omega _{10}{\bf u}_{0}+\omega _{20}{\bf v}_{0},
\qquad
\frac{d{\bf u}_{0}}{ds}=-\omega _{10}{\bf t}_{0}+\omega _{30}{\bf v}_{0},
\qquad
\frac{d{\bf v}_{0}}{ds}=-\omega _{20}{\bf t}_{0}-\omega_{30}{\bf u}_{0},
\label{frenini}
\end{equation}
where $\omega_{k0}(s)$ are generalized intrinsic torsions.
The position of an arbitrary point of the rod is given by the Lagrangian
coordinates $\{\xi _{k}\}$, where $\xi _{1},\xi _{2}$ are Cartesian
coordinates in the cross-sectional plane with unit vectors ${\bf u}_{0}$
and ${\bf v}_{0}$, and $\xi _{3}=s$,
\[ {\bf r}_{0}(\xi _{1},\xi _{2},s)={\bf R}_{0}(s)+\xi _{1}{\bf u}_{0}(s)
+\xi_{2}{\bf v}_{0}(s). \]
We assume that the centerline is inextensible, which means that $s$ remains
the arc-length in the deformed state as well.
The position of the centerline in the deformed configuration
is determined by its radius vector ${\bf R}(s)$.
The position of an arbitrary point of the cross-section with
the arc-length $s$ is
\begin{equation}
{\bf r}(\xi _{1},\xi _{2},s)={\bf R}(s)+\xi_{1}{\bf u}(s)
+\xi _{2}{\bf v}(s),
\label{radfin}
\end{equation}
where ${\bf u}$ and ${\bf v}$ are unit vectors directed along the axes of
symmetry of the cross-section,
and ${\bf t}=d{\bf R}/ds$ is the unit tangent vector.
The functional form of Eq. (\ref{radfin}) is based on the assumption
that after deformation any cross-section remains planar and perpendicular
to the centerline of the rod.
By analogy with Eq. (\ref{frenini}), we write
\begin{equation}
\frac{d{\bf t}}{ds}=\omega _{1}{\bf u}+\omega _{2}{\bf v},
\qquad
\frac{d{\bf u}}{ds}=-\omega _{1}{\bf t}+\omega _{3}{\bf v},
\qquad
\frac{d{\bf v}}{ds}=-\omega _{2}{\bf t}-\omega _{3}{\bf u},
\label{frenet}
\end{equation}
where $\omega _{k}(s)$ are generalized torsions.

Using the methods of Ref. \cite{DR00}, we calculate the displacement
gradient in the deformed state $\partial {\bf r}_{0}/\partial {\bf r}$
and the strain tensor.
The mechanical energy is constructed as a quadratic
form in this strain tensor.
Expressions for the components of the stress tensor
are derived by differentiating this energy with respect
to the corresponding components of the strain tensor.
The constitutive equations relate the components of the internal
moment ${\bf M}=M_{u}{\bf u}+M_{v}{\bf v}+M_{t}{\bf t}$ to
deviations of the generalized torsions from their spontaneous values,
\begin{equation}
M_{u}=-A_{2}(\omega _{2}-\omega_{20}),
\quad
M_{v}=A_{1}(\omega _{1}-\omega_{10}),
\quad
M_{t}=A_{3}(\omega _{3}-\omega _{30}),
\label{constit}
\end{equation}
where $A_{1}=E_{1}I_{1}$, $A_{2}=E_{1}I_{2}$, $A_{3}=E_{2}(I_{1}+I_{2})$,
$I_{1}$ and $I_{2}$ are moments of inertia with respect to the two symmetry
axes of the cross-section, and $E_{1}$ and $E_{2}$ are elastic moduli.
Denote the internal force by
${\bf F}=F_{u}{\bf u}+F_{v}{\bf v}+F_{t}{\bf t}$,
the external distributed force per unit length by
${\bf q}=q_{u}{\bf u}+q_{v}{\bf v}+q_{t}{\bf t}$,
and the external distributed moment per unit length
by ${\bf m}=m_{u}{\bf u}+m_{v}{\bf v}+m_{t}{\bf t}$.
Upon some algebra, we eliminate the components of the
internal force in the cross-sectional plane and
write down the conditions of mechanical equilibrium,
\begin{eqnarray}
&&\frac{dF_{t}}{ds}+\omega _{1}\frac{dM_{v}}{ds}
-\omega_{2}\frac{dM_{u}}{ds}
+\omega _{3}(\omega _{1}M_{u}+\omega _{2}M_{v})
+\omega _{1}m_{v}-\omega _{2}m_{u}+q_{t}=0,
\nonumber\\
&&\frac{dM_{t}}{ds}-\omega _{1}M_{u}-\omega_{2}M_{v}+m_{t}=0,
\nonumber\\
&&\frac{d}{ds}(\frac{dM_{v}}{ds}+\omega_{3}M_{u}
+\omega_{2}M_{t} +m_{v})
+\omega _{3}(\frac{dM_{u}}{ds}-\omega_{3}M_{v}
+\omega _{1}M_{t}+m_{u})
\nonumber\\
&& -\omega _{1}F_{t}-q_{u}=0,
\nonumber\\
&& \frac{d}{ds} (\frac{dM_{u}}{ds}-\omega_{3}M_{v}
+\omega_{1}M_{t} +m_{u})-\omega _{3}(\frac{dM_{v}}{ds}
+\omega_{3}M_{u} +\omega _{2}M_{t}+m_{v})
\nonumber\\
&& +\omega _{2}F_{t}+q_{v}=0.
\label{Munder}
\end{eqnarray}
Given boundary conditions, the four nonlinear differential equations
(\ref{Munder}) together with the constitutive relation
(\ref{constit}), uniquely determine the four unknown functions, $F_{t}$,
$\omega _{1}$, $\omega _{2}$ and $\omega _{3}$.
The generalized torsions $\omega_{k}$ are substituted
into Eqs. (\ref{frenet}) which are resolved to determine
the spatial configuration of the centerline
and the rotation of the cross-section (twist) of the deformed rod.
The tangential force $F_{t}$ plays the role of a Lagrange multiplier
which enforces the inextensibility condition.

\section{Stability of a rectilinear rod}

We begin with the analysis of stability of a spontaneously twisted
rectilinear rod.
In the stress-free reference state, the rod's centerline is straight,
and its cross-sections rotate around the centerline by
some angle $\Omega =\Omega (s)$.
We set $A_{1}<A_{2}$, which means that the vector ${\bf u}_{0}$
coincides with the smaller of the two principal axes of inertia.
The vectors ${\bf u}_{0}$, ${\bf v}_{0}$ and ${\bf t}_{0}$ are given by
${\bf u}_{0}=\cos \Omega \,{\bf e}_{1}+\sin \Omega \,{\bf e}_{2}$,
${\bf v}_{0}=-\sin \Omega \,{\bf e}_{1}+\cos \Omega \,{\bf e}_{2}$,
and ${\bf t}_{0}={\bf e}_{3}$,
where ${\bf e}_{k}$ are unit vectors of some Cartesian coordinate
frame associated with the stress-free rod.
These equalities and Eq. (\ref{frenini}) imply that
\[ \omega _{10}=0,
\qquad
\omega _{20}=0,
\qquad
\omega _{30}=\frac{d\Omega }{ds}. \]
We will assume in the following that $\omega _{30}$ is independent of $s$.

When some compressive force $P$ (smaller than the critical force for
buckling) is applied to the ends of an inextensible rod,
the rod does not deform,
the moments $M_{u}$, $M_{v}$ and $M_{t}$ vanish,
and only a longitudinal internal force $-P$ is generated.
We now assume that, in addition to the load $P$,
small perturbations of external distributed forces
and external distributed moments are present.
These perturbations cause small displacements of the rod
that consist of small displacements of the centerline,
$\Delta {\bf R}=x_{1}{\bf e}_{1}+x_{2}{\bf e}_{2}$,
and small rotations of the cross-sections around the centerline $\alpha $.
Under this perturbation, the vectors
${\bf u}_{0}$, ${\bf v}_{0}$ and ${\bf t}_{0}$
are replaced by ${\bf u}$, ${\bf v}$, ${\bf t}$,
which are substituted into Eq. (\ref{frenet}) to obtain expressions
for the perturbed generalized torsions $\omega _{k}$
(linear in the perturbations $x_{1}$, $x_{2}$ and $\alpha $).
Substitution of these expressions into Eq. (\ref{constit})
yields the perturbations of internal moments,
\[ M_{u}=A_{2}\Bigl(\frac{d^{2}x_{1}}{ds^{2}}\sin \Omega
-\frac{d^{2}x_{2}}{ds^{2}}\cos \Omega \Bigr),
\quad
M_{v}=A_{1}\Bigl(\frac{d^{2}x_{1}}{ds^{2}}\cos \Omega
+\frac{d^{2}x_{2}}{ds^{2}}\sin \Omega \Bigr),
\quad
M_{t}=A_{3}\frac{d\alpha }{ds}. \]
Substituting these expressions into Eqs. (\ref{Munder}),
we find that the equations for the perturbation of the longitudinal
force and for the rotation $\alpha$ decouple from the equations
for the curvatures $\omega _{1}$ and $\omega _{2}$,
which govern bending of the rod and determine it stability
with respect to buckling.

When the external load $P$ reaches its critical value, the rod becomes
unstable against arbitrarily small external perturbations of moments and
forces and, therefore, buckling can be interpreted as the point at which a
non-vanishing solution to Eqs. (\ref{Munder}), in which these forces and
moments are set to zero, first appears.
This yields
\begin{eqnarray}
&&A_{1}\frac{d^{2}\omega _{1}}{ds^{2}}
+(P-A_{1}\omega _{30}^{2})\omega_{1}
-2A_{2}\omega _{30}\frac{d\omega _{2}}{ds}=0,
\nonumber \\
&&A_{2}\frac{d^{2}\omega _{2}}{ds^{2}}
+(P-A_{2}\omega _{30}^{2})\omega_{2}
+2A_{1}\omega _{30}\frac{d\omega _{1}}{ds}=0.
\label{inst1}
\end{eqnarray}
Equations (\ref{inst1}) are supplemented by the geometric relations which
follow from Eqs. (\ref{frenet}):
\[ \frac{d^{2}x_{1}}{ds^{2}}
=\omega _{1}\cos \Omega-\omega _{2}\sin \Omega,
\qquad
\frac{d^{2}x_{2}}{ds^{2}}=\omega _{1}\sin \Omega+\omega _{2}\cos \Omega .\]
The resulting linear equations are solved subject to the boundary
conditions corresponding to hinged ends:
\[ x_{1}(0)=x_{1}(L)=x_{2}(0)=x_{2}(L)=0,
\qquad
\frac{d^{2}x_{1}}{ds^{2}}(0)=\frac{d^{2}x_{1}}{ds^{2}}(L)
=\frac{d^{2}x_{2}}{ds^{2}}(0)=\frac{d^{2}x_{2}}{ds^{2}}(L)=0. \]
When the spontaneous twist $\omega _{30}$ vanishes,
Eqs. (\ref{inst1}) turn into conventional equations for stability
of an elastic beam \cite{TG63}.
In the general case, Eqs. (\ref{inst1}) may be resolved with
respect to $\omega_{1}$ and $\omega _{2}$.
Each of these functions obeys a linear fourth-order
differential equation which contains only even derivatives with respect
to $s$ and can, therefore, be written as a linear combination
of $\ \sin $($\beta_{k}s$) and $\cos (\beta _{k}s)$,
where the ``wavenumbers'' $\beta _{1}$ and $\beta _{2}$
are the positive roots of a characteristic equation that
guarantees the existence of non-vanishing solutions $\omega _{1}(s)$
and $\omega _{2}(s)$ to Eqs. (\ref{inst1}).
Using standard methods, we derive a transcendental equation which
relates the dimensionless critical force
$p^{\ast}=P_{\rm cr}/P_{\rm cr}$,
where $P_{\rm cr}=\pi^{2}A_{1}/L^{2}$ is the Euler critical force
for an untwisted rod \cite{Lov44},
to the total spontaneous twist angle $\Omega (L)=\omega_{30}L$.
The numerical solutions of these equations for various values of
the cross-section asymmetry $\nu =I_{2}/I_{1}$
are plotted in Figure~1.
Note that at the point of the instability,
all generalized torsions $\omega_{k}$ do not vanish,
which implies that, unlike the planar Euler instability,
buckling of a spontaneously twisted rod is intrinsically three-dimensional.

For a circularly symmetric cross-section ($\nu =1$),
the Euler result holds for arbitrary values of the spontaneous twist,
i.e., $p^{\ast }$ is independent of $\Omega $.
For any $\nu >1$, the critical force increases initially
with $\Omega (L)$ and reaches a maximum at $\Omega (L)=2\pi $.
Upon further increase in the total angle of twist,
the critical force oscillates (with additional maxima
at integer multiples of $\pi $) and eventually tends to
some limiting value which depends on $\nu $.
To understand the origin of this non-monotonic behavior,
we investigated the behavior of critical wavenumbers $\beta _{k}$.
Numerical calculations show that the difference between the
critical wavenumbers at buckling, $\beta_{2}-\beta _{1}$, oscillates
as a function of $\Omega (L)$, with maxima occurring at integer
multiples of $\pi $, and tends asymptotically to $2\pi $.
In this asymptotic limit the critical wavenumbers approach
the ``resonance'' conditions
$\beta _{k}L\rightarrow $ $\omega _{30}L \pm \pi $ ($k=1,2$).
We conclude that the oscillatory dependence of the critical
force on the twist angle arises as a result of the interplay
between the two modes.
Whenever the total angle of twist coincides with an integer
multiple of $\pi $, the corresponding axes of symmetry of the
cross--sections at both ends of the rod (at $s=0$ and $s=L$)
become aligned, just as in the twist-free case.

\section{Stability of a ring}

We now turn to study the stability of a circular ring whose centerline
in the stress-free state is a planar circle with radius $R_{0}$.
We introduce cylindrical coordinates $\{r,\phi ,z\}$ with
unit vectors ${\bf e}_{r}$, ${\bf e}_{\phi }$, ${\bf e}_{z}$,
where the axis $z$ is perpendicular to the plane of the ring.
The initial position of the centerline in the stress-free
reference state is determined by the radius vector
${\bf R}_{0}=R_{0}{\bf e}_{r}$.
We assume that in its stress free reference state the rod is
characterized by a twist angle $\Omega =\Omega (s)$ between
the vectors $-{\bf e}_{r}$ and ${\bf u}_{0}$.
The unit vectors ${\bf t}_{0}$, ${\bf u}_{0}$ and ${\bf v}_{0}$
can be expressed in terms of ${\bf e}_{r}$, ${\bf e}_{\phi}$
and ${\bf e}_{z}$ as
\[ {\bf t}_{0}={\bf e}_{\phi },
\qquad
{\bf u}_{0}=-\cos \Omega \,{\bf e}_{r}+\sin \Omega \,{\bf e}_{z},
\qquad
{\bf v}_{0}=\sin \Omega \,{\bf e}_{r}+\cos \Omega \,{\bf e}_{z}. \]
Substitution of these expressions into Eqs. (\ref{frenini}) results in
\[ \omega _{10}=\frac{1}{R_{0}}\cos \Omega,
\qquad
\omega _{20}=-\frac{1}{R_{0}}\sin \Omega,
\qquad
\omega _{30}=\frac{d\Omega }{ds}. \]
We study the stability of the twisted ring under a radially
inward-directed force per unit length ${\bf q}=-q{\bf e}_{r}$
that acts on the outer perimeter of the ring.
As long as the force is smaller than some critical value,
it is balanced by a tangential internal force $F_{t}=-qR_{0}$ and
no displacement arises.
We consider a small deviation of the centerline from
its unperturbed position,
${\bf R}_{0}\rightarrow {\bf R}
={\bf R}_{0}+y_{r}{\bf e}_{r}+y_{\phi }{\bf e}_{\phi }+y_{z}{\bf e}_{z}$,
and a small perturbation of the rotation angle about the centerline,
$\Omega \rightarrow \Omega +\alpha $.
We express the perturbed unit vectors ${\bf t}$, ${\bf u}$ and ${\bf v}$
in terms of the small deviations
$y_{\phi }$, $y_{z}$ and $\alpha $ ($y_{r}$ is eliminated
using the inextensibility condition $|{\bf t}|=1$)
and use Eq. (\ref{frenet}) to relate these deviations to
the perturbed spontaneous torsions $\omega _{k}$.
The subsequent analysis is qualitatively similar
(but much more cumbersome) to that leading to Eqs. (\ref{inst1}).
We derive three coupled differential equations with variable coefficients
for the functions $y_{\phi }(s)$, $y_{z}(s)$ and $\alpha(s)$.
An analytical solution of these equations can be obtained only
in the special case of equal moments of inertia,
$I_{1}=I_{2}$ (e.g., for a circular cross-section).
In this case we find that buckling takes place when the
critical force per unit length reaches a critical value
\[ q_{\rm cr}=\frac{3E_{1}I_{1}}{R_{0}^{3}}. \]
As expected for a circular rod, the critical force does not depend on the
spontaneous twist $\Omega $.
In this case, the onset of instability with respect to in-plane buckling
of a circular rod coincides with that for out-of-plane buckling.

In the general case, $I_{1}<I_{2}$, we have to resort to approximate
methods.
For simplicity, we set $\Omega(s)=Ns/R_{0}$, where $N$ is a positive
integer.
A reasonable estimate which is expected to capture the qualitative
aspects of the effect of spontaneous twist on the critical force,
can be obtained using the Galerkin method \cite{TG63}.
The perturbations $y_{\phi }$, $y_{z}$ and $\alpha $ are presented
in the form $X_{n}\sin (ns/R_{0})$ with a fixed $n$.
We substitute these expressions into the equilibrium equations,
multiply both sides by $\sin (ns/R_{0})$ and require that
the difference between the left and the right hand sides of
the resulting equalities vanishes on the average
(i.e., when integrated over the ring contour,
from $s=0$ to $2\pi R_{0}$).
Upon some algebra, we arrive at the following expression for the critical
compressive force per unit length
\begin{equation}
q_{\rm cr}=\frac{E_{2}(I_{1}+I_{2})}{2R_{0}^{3}}\biggl [
3-\frac{I_{2}-I_{1}}{I_{2}+I_{1}}(N^{2}+1)\biggr ].
\label{crit2}
\end{equation}
Note that for $N^{2}\geq 3(I_{1}+I_{2})/(I_{2}-I_{1})-1$ the ring becomes
unstable with respect to buckling, even in the absence of external forces.
Because the equations for the perturbations are coupled,
the instability has a three dimensional character
and buckling is not confined to the plane of the ring.

It is of interest to compare these conclusions with the results of
Ref. \cite{GT97}, where the stability of the stress-free configuration
of a circular ring was analyzed and it was found that, in general,
there were no non-trivial solutions to the equilibrium equations
with periodic boundary conditions.
The apparent discrepancy stems from the fact that the problems
under consideration are rather different:
while we examine the stability of a ring with spontaneous twist
under compression in the plane of the ring,
Ref. \cite{GT97} analyzes the stability of a ring
without spontaneous twist under the action of an external torque
applied to a fixed cross-section.

\section{Conclusions}
Buckling instabilities have been studied for thin rods with
spontaneous twist and non-circular cross-sections.
Two cases are analyzed in detail: (i) a rectilinear rod compressed
by forces applied to its ends,
and (ii) a circular ring compressed by uniformly distributed pressure
acting in its plane.
It is demonstrated that for a rectilinear rod,
an increase in the spontaneous twist angle leads to
the growth of the critical force and stabilizes the rod against buckling.
The effect increases with the ratio of principal moments
of inertia, $\nu =I_{2}/I_{1}>1$.
For a circular ring, an increase in the spontaneous twist leads
to a monotonic decrease in the critical force.
The rate of the decrease grows with $\nu$.
For sufficiently large values of the spontaneous twist,
the ring may become unstable to buckling even in the absence
of external forces.

\subsection*{Acknowledgement}
We would like to thank S. Panyukov for helpful comments.
AD gratefully acknowledges financial support by the Israeli Ministry of
Science through grant 1202--1--98.
YR's work is supported by a grant from the Israel Science Foundation.

\newpage

\setlength{\unitlength}{0.8 mm}
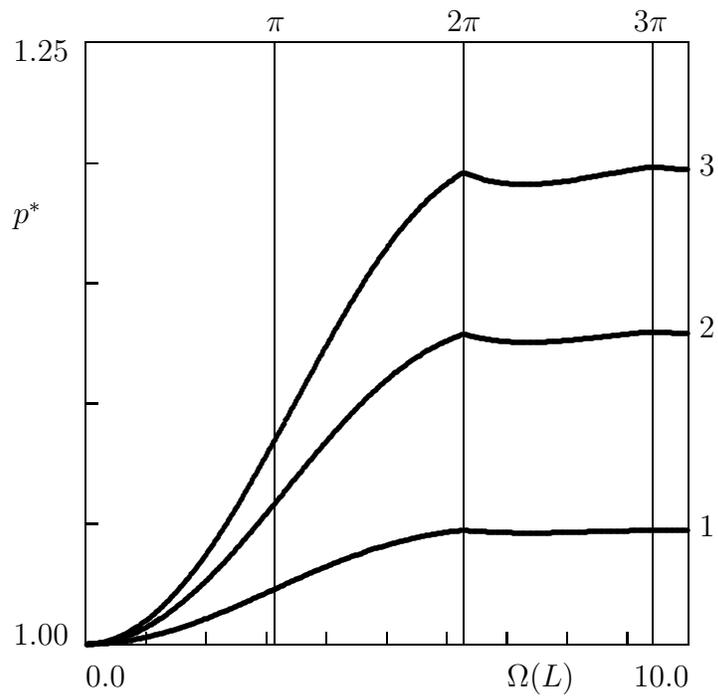
\begin{figure}[t]
\begin{center}
\begin{picture}(100,100)
\put(0,0){\framebox(100,100)}
\multiput(10,0)(10,0){9}{\line(0,1){2}}
\multiput(0,20)(0,20){4}{\line(1,0){2}}
\put(-12,0){1.00}
\put(-12,97){1.25}
\put(-12,70){$p^{\ast}$}
\put(0, -7){0.0}
\put(91,-7){10.0}
\put(70,-7){$\Omega(L)$}

\put(102,18){1}
\put(102,51){2}
\put(102,78){3}
\put(31.4, 0){\line(0,1){100}}
\put(62.8, 0){\line(0,1){100}}
\put(94.2, 0){\line(0,1){100}}
\put(30,102){$\pi$}
\put(60,102){$2\pi$}
\put(91,102){$3\pi$}

\put(   0.25,    0.00){\circle*{1.0}} 
\put(   0.50,    0.01){\circle*{1.0}} 
\put(   0.75,    0.02){\circle*{1.0}} 
\put(   1.00,    0.02){\circle*{1.0}} 
\put(   1.25,    0.02){\circle*{1.0}} 
\put(   1.50,    0.05){\circle*{1.0}} 
\put(   1.75,    0.04){\circle*{1.0}} 
\put(   2.00,    0.06){\circle*{1.0}} 
\put(   2.25,    0.08){\circle*{1.0}} 
\put(   2.50,    0.09){\circle*{1.0}} 
\put(   2.75,    0.10){\circle*{1.0}} 
\put(   3.00,    0.14){\circle*{1.0}} 
\put(   3.25,    0.15){\circle*{1.0}} 
\put(   3.50,    0.17){\circle*{1.0}} 
\put(   3.75,    0.17){\circle*{1.0}} 
\put(   4.00,    0.22){\circle*{1.0}} 
\put(   4.25,    0.25){\circle*{1.0}} 
\put(   4.50,    0.25){\circle*{1.0}} 
\put(   4.75,    0.27){\circle*{1.0}} 
\put(   5.00,    0.30){\circle*{1.0}} 
\put(   5.25,    0.35){\circle*{1.0}} 
\put(   5.50,    0.37){\circle*{1.0}} 
\put(   5.75,    0.41){\circle*{1.0}} 
\put(   6.00,    0.42){\circle*{1.0}} 
\put(   6.25,    0.45){\circle*{1.0}} 
\put(   6.50,    0.49){\circle*{1.0}} 
\put(   6.75,    0.55){\circle*{1.0}} 
\put(   7.00,    0.58){\circle*{1.0}} 
\put(   7.25,    0.63){\circle*{1.0}} 
\put(   7.50,    0.65){\circle*{1.0}} 
\put(   7.75,    0.69){\circle*{1.0}} 
\put(   8.00,    0.75){\circle*{1.0}} 
\put(   8.25,    0.81){\circle*{1.0}} 
\put(   8.50,    0.86){\circle*{1.0}} 
\put(   8.75,    0.87){\circle*{1.0}} 
\put(   9.00,    0.95){\circle*{1.0}} 
\put(   9.25,    1.00){\circle*{1.0}} 
\put(   9.50,    1.06){\circle*{1.0}} 
\put(   9.75,    1.10){\circle*{1.0}} 
\put(  10.00,    1.15){\circle*{1.0}} 
\put(  10.25,    1.22){\circle*{1.0}} 
\put(  10.50,    1.26){\circle*{1.0}} 
\put(  10.75,    1.32){\circle*{1.0}} 
\put(  11.00,    1.39){\circle*{1.0}} 
\put(  11.25,    1.44){\circle*{1.0}} 
\put(  11.50,    1.51){\circle*{1.0}} 
\put(  11.75,    1.58){\circle*{1.0}} 
\put(  12.00,    1.64){\circle*{1.0}} 
\put(  12.25,    1.71){\circle*{1.0}} 
\put(  12.50,    1.75){\circle*{1.0}} 
\put(  12.75,    1.85){\circle*{1.0}} 
\put(  13.00,    1.88){\circle*{1.0}} 
\put(  13.25,    1.97){\circle*{1.0}} 
\put(  13.50,    2.03){\circle*{1.0}} 
\put(  13.75,    2.11){\circle*{1.0}} 
\put(  14.00,    2.20){\circle*{1.0}} 
\put(  14.25,    2.27){\circle*{1.0}} 
\put(  14.50,    2.32){\circle*{1.0}} 
\put(  14.75,    2.41){\circle*{1.0}} 
\put(  15.00,    2.49){\circle*{1.0}} 
\put(  15.25,    2.58){\circle*{1.0}} 
\put(  15.50,    2.64){\circle*{1.0}} 
\put(  15.75,    2.72){\circle*{1.0}} 
\put(  16.00,    2.81){\circle*{1.0}} 
\put(  16.25,    2.88){\circle*{1.0}} 
\put(  16.50,    2.96){\circle*{1.0}} 
\put(  16.75,    3.06){\circle*{1.0}} 
\put(  17.00,    3.14){\circle*{1.0}} 
\put(  17.25,    3.23){\circle*{1.0}} 
\put(  17.50,    3.33){\circle*{1.0}} 
\put(  17.75,    3.41){\circle*{1.0}} 
\put(  18.00,    3.50){\circle*{1.0}} 
\put(  18.25,    3.61){\circle*{1.0}} 
\put(  18.50,    3.69){\circle*{1.0}} 
\put(  18.75,    3.79){\circle*{1.0}} 
\put(  19.00,    3.87){\circle*{1.0}} 
\put(  19.25,    3.95){\circle*{1.0}} 
\put(  19.50,    4.06){\circle*{1.0}} 
\put(  19.75,    4.14){\circle*{1.0}} 
\put(  20.00,    4.23){\circle*{1.0}} 
\put(  20.25,    4.34){\circle*{1.0}} 
\put(  20.50,    4.42){\circle*{1.0}} 
\put(  20.75,    4.52){\circle*{1.0}} 
\put(  21.00,    4.64){\circle*{1.0}} 
\put(  21.25,    4.72){\circle*{1.0}} 
\put(  21.50,    4.83){\circle*{1.0}} 
\put(  21.75,    4.95){\circle*{1.0}} 
\put(  22.00,    5.04){\circle*{1.0}} 
\put(  22.25,    5.15){\circle*{1.0}} 
\put(  22.50,    5.23){\circle*{1.0}} 
\put(  22.75,    5.33){\circle*{1.0}} 
\put(  23.00,    5.45){\circle*{1.0}} 
\put(  23.25,    5.54){\circle*{1.0}} 
\put(  23.50,    5.64){\circle*{1.0}} 
\put(  23.75,    5.76){\circle*{1.0}} 
\put(  24.00,    5.85){\circle*{1.0}} 
\put(  24.25,    5.96){\circle*{1.0}} 
\put(  24.50,    6.09){\circle*{1.0}} 
\put(  24.75,    6.18){\circle*{1.0}} 
\put(  25.00,    6.30){\circle*{1.0}} 
\put(  25.25,    6.39){\circle*{1.0}} 
\put(  25.50,    6.53){\circle*{1.0}} 
\put(  25.75,    6.61){\circle*{1.0}} 
\put(  26.00,    6.74){\circle*{1.0}} 
\put(  26.25,    6.85){\circle*{1.0}} 
\put(  26.50,    6.94){\circle*{1.0}} 
\put(  26.75,    7.08){\circle*{1.0}} 
\put(  27.00,    7.19){\circle*{1.0}} 
\put(  27.25,    7.28){\circle*{1.0}} 
\put(  27.50,    7.38){\circle*{1.0}} 
\put(  27.75,    7.50){\circle*{1.0}} 
\put(  28.00,    7.64){\circle*{1.0}} 
\put(  28.25,    7.74){\circle*{1.0}} 
\put(  28.50,    7.83){\circle*{1.0}} 
\put(  28.75,    7.97){\circle*{1.0}} 
\put(  29.00,    8.08){\circle*{1.0}} 
\put(  29.25,    8.17){\circle*{1.0}} 
\put(  29.50,    8.27){\circle*{1.0}} 
\put(  29.75,    8.39){\circle*{1.0}} 
\put(  30.00,    8.53){\circle*{1.0}} 
\put(  30.25,    8.64){\circle*{1.0}}

\put(  30.43,    8.72){\circle*{1.0}}
\put(  30.60,    8.80){\circle*{1.0}}
\put(  30.78,    8.87){\circle*{1.0}}
\put(  30.95,    8.95){\circle*{1.0}}
\put(  31.12,    9.03){\circle*{1.0}}
\put(  31.30,    9.11){\circle*{1.0}}
\put(  31.48,    9.19){\circle*{1.0}}
\put(  31.65,    9.26){\circle*{1.0}}
\put(  31.83,    9.34){\circle*{1.0}}

\put(  32.00,    9.42){\circle*{1.0}} 
\put(  32.25,    9.58){\circle*{1.0}} 
\put(  32.50,    9.66){\circle*{1.0}} 
\put(  32.75,    9.74){\circle*{1.0}} 
\put(  33.00,    9.90){\circle*{1.0}} 
\put(  33.25,    9.99){\circle*{1.0}} 
\put(  33.50,   10.15){\circle*{1.0}} 
\put(  33.75,   10.23){\circle*{1.0}} 
\put(  34.00,   10.31){\circle*{1.0}} 
\put(  34.25,   10.47){\circle*{1.0}} 
\put(  34.50,   10.55){\circle*{1.0}} 
\put(  34.75,   10.72){\circle*{1.0}} 
\put(  35.00,   10.80){\circle*{1.0}} 
\put(  35.25,   10.88){\circle*{1.0}} 
\put(  35.50,   11.04){\circle*{1.0}} 
\put(  35.75,   11.12){\circle*{1.0}} 
\put(  36.00,   11.20){\circle*{1.0}} 
\put(  36.25,   11.36){\circle*{1.0}} 
\put(  36.50,   11.45){\circle*{1.0}} 
\put(  36.75,   11.53){\circle*{1.0}} 
\put(  37.00,   11.69){\circle*{1.0}} 
\put(  37.25,   11.77){\circle*{1.0}} 
\put(  37.50,   11.85){\circle*{1.0}} 
\put(  37.75,   12.01){\circle*{1.0}} 
\put(  38.00,   12.09){\circle*{1.0}} 
\put(  38.25,   12.17){\circle*{1.0}} 
\put(  38.50,   12.34){\circle*{1.0}} 
\put(  38.75,   12.42){\circle*{1.0}} 
\put(  39.00,   12.50){\circle*{1.0}} 
\put(  39.25,   12.66){\circle*{1.0}} 
\put(  39.50,   12.74){\circle*{1.0}} 
\put(  39.75,   12.82){\circle*{1.0}} 
\put(  40.00,   12.90){\circle*{1.0}} 
\put(  40.25,   13.07){\circle*{1.0}} 
\put(  40.50,   13.15){\circle*{1.0}} 
\put(  40.75,   13.23){\circle*{1.0}} 
\put(  41.00,   13.31){\circle*{1.0}} 
\put(  41.25,   13.47){\circle*{1.0}} 
\put(  41.50,   13.55){\circle*{1.0}} 
\put(  41.75,   13.63){\circle*{1.0}} 
\put(  42.00,   13.71){\circle*{1.0}} 
\put(  42.25,   13.80){\circle*{1.0}} 
\put(  42.50,   13.96){\circle*{1.0}} 
\put(  42.75,   14.04){\circle*{1.0}} 
\put(  43.00,   14.12){\circle*{1.0}} 
\put(  43.25,   14.20){\circle*{1.0}} 
\put(  43.50,   14.28){\circle*{1.0}} 
\put(  43.75,   14.36){\circle*{1.0}} 
\put(  44.00,   14.53){\circle*{1.0}} 
\put(  44.25,   14.61){\circle*{1.0}} 
\put(  44.50,   14.69){\circle*{1.0}} 
\put(  44.75,   14.77){\circle*{1.0}} 
\put(  45.00,   14.85){\circle*{1.0}} 
\put(  45.25,   14.93){\circle*{1.0}} 
\put(  45.50,   15.01){\circle*{1.0}} 
\put(  45.75,   15.09){\circle*{1.0}} 
\put(  46.00,   15.17){\circle*{1.0}} 
\put(  46.25,   15.25){\circle*{1.0}} 
\put(  46.50,   15.34){\circle*{1.0}} 
\put(  46.75,   15.50){\circle*{1.0}} 
\put(  47.00,   15.58){\circle*{1.0}} 
\put(  47.25,   15.66){\circle*{1.0}} 
\put(  47.50,   15.74){\circle*{1.0}} 
\put(  47.75,   15.82){\circle*{1.0}} 
\put(  48.00,   15.90){\circle*{1.0}} 
\put(  48.25,   15.98){\circle*{1.0}} 
\put(  48.50,   15.98){\circle*{1.0}} 
\put(  48.75,   16.07){\circle*{1.0}} 
\put(  49.00,   16.15){\circle*{1.0}} 
\put(  49.25,   16.23){\circle*{1.0}} 
\put(  49.50,   16.31){\circle*{1.0}} 
\put(  49.75,   16.39){\circle*{1.0}} 
\put(  50.00,   16.47){\circle*{1.0}} 
\put(  50.25,   16.55){\circle*{1.0}} 
\put(  50.50,   16.63){\circle*{1.0}} 
\put(  50.75,   16.71){\circle*{1.0}} 
\put(  51.00,   16.79){\circle*{1.0}} 
\put(  51.25,   16.79){\circle*{1.0}} 
\put(  51.50,   16.88){\circle*{1.0}} 
\put(  51.75,   16.96){\circle*{1.0}} 
\put(  52.00,   17.04){\circle*{1.0}} 
\put(  52.25,   17.12){\circle*{1.0}} 
\put(  52.50,   17.12){\circle*{1.0}} 
\put(  52.75,   17.20){\circle*{1.0}} 
\put(  53.00,   17.28){\circle*{1.0}} 
\put(  53.25,   17.36){\circle*{1.0}} 
\put(  53.50,   17.36){\circle*{1.0}} 
\put(  53.75,   17.44){\circle*{1.0}} 
\put(  54.00,   17.52){\circle*{1.0}} 
\put(  54.25,   17.61){\circle*{1.0}} 
\put(  54.50,   17.61){\circle*{1.0}} 
\put(  54.75,   17.69){\circle*{1.0}} 
\put(  55.00,   17.77){\circle*{1.0}} 
\put(  55.25,   17.77){\circle*{1.0}} 
\put(  55.50,   17.85){\circle*{1.0}} 
\put(  55.75,   17.93){\circle*{1.0}} 
\put(  56.00,   17.93){\circle*{1.0}} 
\put(  56.25,   18.01){\circle*{1.0}} 
\put(  56.50,   18.09){\circle*{1.0}} 
\put(  56.75,   18.09){\circle*{1.0}} 
\put(  57.00,   18.17){\circle*{1.0}} 
\put(  57.25,   18.17){\circle*{1.0}} 
\put(  57.50,   18.25){\circle*{1.0}} 
\put(  57.75,   18.25){\circle*{1.0}} 
\put(  58.00,   18.33){\circle*{1.0}} 
\put(  58.25,   18.33){\circle*{1.0}} 
\put(  58.50,   18.42){\circle*{1.0}} 
\put(  58.75,   18.42){\circle*{1.0}} 
\put(  59.00,   18.50){\circle*{1.0}} 
\put(  59.25,   18.50){\circle*{1.0}} 
\put(  59.50,   18.58){\circle*{1.0}} 
\put(  59.75,   18.58){\circle*{1.0}} 
\put(  60.00,   18.66){\circle*{1.0}} 
\put(  60.25,   18.66){\circle*{1.0}} 
\put(  60.50,   18.74){\circle*{1.0}} 
\put(  60.75,   18.74){\circle*{1.0}} 
\put(  61.00,   18.82){\circle*{1.0}} 
\put(  61.25,   18.82){\circle*{1.0}} 
\put(  61.50,   18.82){\circle*{1.0}} 
\put(  61.75,   18.90){\circle*{1.0}} 
\put(  62.00,   18.90){\circle*{1.0}} 
\put(  62.25,   18.98){\circle*{1.0}} 
\put(  62.50,   18.98){\circle*{1.0}} 
\put(  63.00,   18.98){\circle*{1.0}} 
\put(  63.25,   18.98){\circle*{1.0}} 
\put(  63.50,   18.98){\circle*{1.0}} 
\put(  63.75,   18.90){\circle*{1.0}} 
\put(  64.00,   18.90){\circle*{1.0}} 
\put(  64.25,   18.90){\circle*{1.0}} 
\put(  64.50,   18.82){\circle*{1.0}} 
\put(  64.75,   18.82){\circle*{1.0}} 
\put(  65.00,   18.82){\circle*{1.0}} 
\put(  65.25,   18.82){\circle*{1.0}} 
\put(  65.50,   18.74){\circle*{1.0}} 
\put(  65.75,   18.74){\circle*{1.0}} 
\put(  66.00,   18.74){\circle*{1.0}} 
\put(  66.25,   18.74){\circle*{1.0}} 
\put(  66.50,   18.74){\circle*{1.0}} 
\put(  66.75,   18.66){\circle*{1.0}} 
\put(  67.00,   18.66){\circle*{1.0}} 
\put(  67.25,   18.66){\circle*{1.0}} 
\put(  67.50,   18.66){\circle*{1.0}} 
\put(  67.75,   18.66){\circle*{1.0}} 
\put(  68.00,   18.66){\circle*{1.0}} 
\put(  68.25,   18.66){\circle*{1.0}} 
\put(  68.50,   18.58){\circle*{1.0}} 
\put(  68.75,   18.58){\circle*{1.0}} 
\put(  69.00,   18.58){\circle*{1.0}} 
\put(  69.25,   18.58){\circle*{1.0}} 
\put(  69.50,   18.58){\circle*{1.0}} 
\put(  69.75,   18.58){\circle*{1.0}} 
\put(  70.00,   18.58){\circle*{1.0}} 
\put(  70.25,   18.58){\circle*{1.0}} 
\put(  70.50,   18.58){\circle*{1.0}} 
\put(  70.75,   18.58){\circle*{1.0}} 
\put(  71.00,   18.50){\circle*{1.0}} 
\put(  71.25,   18.50){\circle*{1.0}} 
\put(  71.50,   18.50){\circle*{1.0}} 
\put(  71.75,   18.50){\circle*{1.0}} 
\put(  72.00,   18.50){\circle*{1.0}} 
\put(  72.25,   18.50){\circle*{1.0}} 
\put(  72.50,   18.50){\circle*{1.0}} 
\put(  72.75,   18.50){\circle*{1.0}} 
\put(  73.00,   18.50){\circle*{1.0}} 
\put(  73.25,   18.50){\circle*{1.0}} 
\put(  73.50,   18.50){\circle*{1.0}} 
\put(  73.75,   18.50){\circle*{1.0}} 
\put(  74.00,   18.50){\circle*{1.0}} 
\put(  74.25,   18.50){\circle*{1.0}} 
\put(  74.50,   18.50){\circle*{1.0}} 
\put(  74.75,   18.50){\circle*{1.0}} 
\put(  75.00,   18.50){\circle*{1.0}} 
\put(  75.25,   18.50){\circle*{1.0}} 
\put(  75.50,   18.50){\circle*{1.0}} 
\put(  75.75,   18.50){\circle*{1.0}} 
\put(  76.00,   18.50){\circle*{1.0}} 
\put(  76.25,   18.50){\circle*{1.0}} 
\put(  76.50,   18.50){\circle*{1.0}} 
\put(  76.75,   18.50){\circle*{1.0}} 
\put(  77.00,   18.50){\circle*{1.0}} 
\put(  77.25,   18.50){\circle*{1.0}} 
\put(  77.50,   18.58){\circle*{1.0}} 
\put(  77.75,   18.58){\circle*{1.0}} 
\put(  78.00,   18.58){\circle*{1.0}} 
\put(  78.25,   18.58){\circle*{1.0}} 
\put(  78.50,   18.58){\circle*{1.0}} 
\put(  78.75,   18.58){\circle*{1.0}} 
\put(  79.00,   18.58){\circle*{1.0}} 
\put(  79.25,   18.58){\circle*{1.0}} 
\put(  79.50,   18.58){\circle*{1.0}} 
\put(  79.75,   18.58){\circle*{1.0}} 
\put(  80.00,   18.58){\circle*{1.0}} 
\put(  80.25,   18.58){\circle*{1.0}} 
\put(  80.50,   18.58){\circle*{1.0}} 
\put(  80.75,   18.58){\circle*{1.0}} 
\put(  81.00,   18.66){\circle*{1.0}} 
\put(  81.25,   18.66){\circle*{1.0}} 
\put(  81.50,   18.66){\circle*{1.0}} 
\put(  81.75,   18.66){\circle*{1.0}} 
\put(  82.00,   18.66){\circle*{1.0}} 
\put(  82.25,   18.66){\circle*{1.0}} 
\put(  82.50,   18.66){\circle*{1.0}} 
\put(  82.75,   18.66){\circle*{1.0}} 
\put(  83.00,   18.66){\circle*{1.0}} 
\put(  83.25,   18.66){\circle*{1.0}} 
\put(  83.50,   18.74){\circle*{1.0}} 
\put(  83.75,   18.74){\circle*{1.0}} 
\put(  84.00,   18.74){\circle*{1.0}} 
\put(  84.25,   18.74){\circle*{1.0}} 
\put(  84.50,   18.74){\circle*{1.0}} 
\put(  84.75,   18.74){\circle*{1.0}} 
\put(  85.00,   18.74){\circle*{1.0}} 
\put(  85.25,   18.74){\circle*{1.0}} 
\put(  85.50,   18.74){\circle*{1.0}} 
\put(  85.75,   18.82){\circle*{1.0}} 
\put(  86.00,   18.82){\circle*{1.0}} 
\put(  86.25,   18.82){\circle*{1.0}} 
\put(  86.50,   18.82){\circle*{1.0}} 
\put(  86.75,   18.82){\circle*{1.0}} 
\put(  87.00,   18.82){\circle*{1.0}} 
\put(  87.25,   18.82){\circle*{1.0}} 
\put(  87.50,   18.82){\circle*{1.0}} 
\put(  87.75,   18.82){\circle*{1.0}} 
\put(  88.00,   18.82){\circle*{1.0}} 
\put(  88.25,   18.90){\circle*{1.0}} 
\put(  88.50,   18.90){\circle*{1.0}} 
\put(  88.75,   18.90){\circle*{1.0}} 
\put(  89.00,   18.90){\circle*{1.0}} 
\put(  89.25,   18.90){\circle*{1.0}} 
\put(  89.50,   18.90){\circle*{1.0}} 
\put(  89.75,   18.90){\circle*{1.0}} 
\put(  90.00,   18.90){\circle*{1.0}} 
\put(  90.25,   18.90){\circle*{1.0}} 
\put(  90.50,   18.98){\circle*{1.0}} 
\put(  90.75,   18.98){\circle*{1.0}} 
\put(  91.00,   18.98){\circle*{1.0}} 
\put(  91.25,   18.98){\circle*{1.0}} 
\put(  91.50,   18.98){\circle*{1.0}} 
\put(  91.75,   18.98){\circle*{1.0}} 
\put(  92.00,   18.98){\circle*{1.0}} 
\put(  92.25,   18.98){\circle*{1.0}} 
\put(  92.50,   18.98){\circle*{1.0}} 
\put(  92.75,   18.98){\circle*{1.0}} 
\put(  93.00,   18.98){\circle*{1.0}}
\put(  93.25,   18.98){\circle*{1.0}}
\put(  93.50,   18.98){\circle*{1.0}}
\put(  93.75,   18.98){\circle*{1.0}}
\put(  94.00,   18.98){\circle*{1.0}}
\put(  94.25,   18.98){\circle*{1.0}}
\put(  94.50,   18.98){\circle*{1.0}}
\put(  94.75,   18.98){\circle*{1.0}}
\put(  95.00,   18.98){\circle*{1.0}}
\put(  95.25,   18.98){\circle*{1.0}}
\put(  95.50,   18.98){\circle*{1.0}}
\put(  95.75,   18.98){\circle*{1.0}}
\put(  96.00,   18.98){\circle*{1.0}} 
\put(  96.25,   18.98){\circle*{1.0}} 
\put(  96.50,   18.98){\circle*{1.0}} 
\put(  96.75,   18.98){\circle*{1.0}} 
\put(  97.00,   18.98){\circle*{1.0}} 
\put(  97.25,   18.98){\circle*{1.0}} 
\put(  97.50,   18.98){\circle*{1.0}} 
\put(  97.75,   18.98){\circle*{1.0}} 
\put(  98.00,   18.98){\circle*{1.0}} 
\put(  98.25,   18.98){\circle*{1.0}} 
\put(  98.50,   18.98){\circle*{1.0}} 
\put(  98.75,   18.98){\circle*{1.0}} 
\put(  99.00,   18.98){\circle*{1.0}} 
\put(  99.25,   18.98){\circle*{1.0}} 
\put(  99.50,   18.98){\circle*{1.0}} 
\put(  99.75,   18.98){\circle*{1.0}} 
\put( 100.00,   18.98){\circle*{1.0}} 

\put(   0.25,    0.01){\circle*{1.0}}
\put(   0.50,    0.03){\circle*{1.0}} 
\put(   0.75,    0.03){\circle*{1.0}} 
\put(   1.00,    0.06){\circle*{1.0}} 
\put(   1.25,    0.07){\circle*{1.0}} 
\put(   1.50,    0.07){\circle*{1.0}} 
\put(   1.75,    0.09){\circle*{1.0}} 
\put(   2.00,    0.14){\circle*{1.0}} 
\put(   2.25,    0.17){\circle*{1.0}} 
\put(   2.50,    0.19){\circle*{1.0}} 
\put(   2.75,    0.23){\circle*{1.0}} 
\put(   3.00,    0.26){\circle*{1.0}} 
\put(   3.25,    0.31){\circle*{1.0}} 
\put(   3.50,    0.35){\circle*{1.0}} 
\put(   3.75,    0.41){\circle*{1.0}} 
\put(   4.00,    0.46){\circle*{1.0}} 
\put(   4.25,    0.54){\circle*{1.0}} 
\put(   4.50,    0.59){\circle*{1.0}} 
\put(   4.75,    0.64){\circle*{1.0}} 
\put(   5.00,    0.71){\circle*{1.0}} 
\put(   5.25,    0.80){\circle*{1.0}} 
\put(   5.50,    0.88){\circle*{1.0}} 
\put(   5.75,    0.94){\circle*{1.0}} 
\put(   6.00,    1.03){\circle*{1.0}} 
\put(   6.25,    1.10){\circle*{1.0}} 
\put(   6.50,    1.20){\circle*{1.0}} 
\put(   6.75,    1.29){\circle*{1.0}} 
\put(   7.00,    1.39){\circle*{1.0}} 
\put(   7.25,    1.49){\circle*{1.0}} 
\put(   7.50,    1.57){\circle*{1.0}} 
\put(   7.75,    1.67){\circle*{1.0}} 
\put(   8.00,    1.80){\circle*{1.0}} 
\put(   8.25,    1.91){\circle*{1.0}} 
\put(   8.50,    2.01){\circle*{1.0}} 
\put(   8.75,    2.14){\circle*{1.0}} 
\put(   9.00,    2.25){\circle*{1.0}} 
\put(   9.25,    2.38){\circle*{1.0}} 
\put(   9.50,    2.50){\circle*{1.0}} 
\put(   9.75,    2.64){\circle*{1.0}} 
\put(  10.00,    2.77){\circle*{1.0}} 
\put(  10.25,    2.93){\circle*{1.0}} 
\put(  10.50,    3.07){\circle*{1.0}} 
\put(  10.75,    3.19){\circle*{1.0}} 
\put(  11.00,    3.34){\circle*{1.0}} 
\put(  11.25,    3.51){\circle*{1.0}} 
\put(  11.50,    3.63){\circle*{1.0}} 
\put(  11.75,    3.82){\circle*{1.0}} 
\put(  12.00,    3.95){\circle*{1.0}} 
\put(  12.25,    4.14){\circle*{1.0}} 
\put(  12.50,    4.28){\circle*{1.0}} 
\put(  12.75,    4.45){\circle*{1.0}} 
\put(  13.00,    4.64){\circle*{1.0}} 
\put(  13.25,    4.81){\circle*{1.0}} 
\put(  13.50,    4.97){\circle*{1.0}} 
\put(  13.75,    5.16){\circle*{1.0}} 
\put(  14.00,    5.33){\circle*{1.0}} 
\put(  14.25,    5.52){\circle*{1.0}} 
\put(  14.50,    5.70){\circle*{1.0}} 
\put(  14.75,    5.91){\circle*{1.0}} 
\put(  15.00,    6.10){\circle*{1.0}} 
\put(  15.25,    6.31){\circle*{1.0}} 
\put(  15.50,    6.51){\circle*{1.0}} 
\put(  15.75,    6.70){\circle*{1.0}} 
\put(  16.00,    6.91){\circle*{1.0}} 
\put(  16.25,    7.10){\circle*{1.0}} 
\put(  16.50,    7.32){\circle*{1.0}} 
\put(  16.75,    7.53){\circle*{1.0}} 
\put(  17.00,    7.76){\circle*{1.0}} 
\put(  17.25,    7.97){\circle*{1.0}} 
\put(  17.50,    8.17){\circle*{1.0}} 
\put(  17.75,    8.40){\circle*{1.0}} 
\put(  18.00,    8.65){\circle*{1.0}} 
\put(  18.25,    8.84){\circle*{1.0}} 
\put(  18.50,    9.10){\circle*{1.0}} 
\put(  18.75,    9.31){\circle*{1.0}} 
\put(  19.00,    9.54){\circle*{1.0}} 
\put(  19.25,    9.80){\circle*{1.0}} 
\put(  19.50,   10.04){\circle*{1.0}} 
\put(  19.75,   10.26){\circle*{1.0}} 
\put(  20.00,   10.51){\circle*{1.0}} 
\put(  20.25,   10.75){\circle*{1.0}} 
\put(  20.50,   11.01){\circle*{1.0}} 
\put(  20.75,   11.26){\circle*{1.0}} 
\put(  21.00,   11.53){\circle*{1.0}} 
\put(  21.25,   11.78){\circle*{1.0}} 
\put(  21.50,   12.02){\circle*{1.0}} 
\put(  21.75,   12.29){\circle*{1.0}} 
\put(  22.00,   12.54){\circle*{1.0}} 
\put(  22.25,   12.82){\circle*{1.0}} 
\put(  22.50,   13.08){\circle*{1.0}} 
\put(  22.75,   13.32){\circle*{1.0}} 
\put(  23.00,   13.59){\circle*{1.0}} 
\put(  23.25,   13.85){\circle*{1.0}} 
\put(  23.50,   14.13){\circle*{1.0}} 
\put(  23.75,   14.40){\circle*{1.0}} 
\put(  24.00,   14.69){\circle*{1.0}} 
\put(  24.25,   14.96){\circle*{1.0}} 
\put(  24.50,   15.22){\circle*{1.0}} 
\put(  24.75,   15.51){\circle*{1.0}} 
\put(  25.00,   15.78){\circle*{1.0}} 
\put(  25.25,   16.08){\circle*{1.0}} 
\put(  25.50,   16.36){\circle*{1.0}} 
\put(  25.75,   16.63){\circle*{1.0}} 
\put(  26.00,   16.92){\circle*{1.0}} 
\put(  26.25,   17.23){\circle*{1.0}} 
\put(  26.50,   17.49){\circle*{1.0}} 
\put(  26.75,   17.78){\circle*{1.0}} 
\put(  27.00,   18.09){\circle*{1.0}} 
\put(  27.25,   18.39){\circle*{1.0}} 
\put(  27.50,   18.67){\circle*{1.0}} 
\put(  27.75,   18.98){\circle*{1.0}} 
\put(  28.00,   19.27){\circle*{1.0}}
 
\put(  28.25,   19.57){\circle*{1.0}}
\put(  28.50,   19.87){\circle*{1.0}}
\put(  28.75,   20.17){\circle*{1.0}}
\put(  29.00,   20.47){\circle*{1.0}}
\put(  29.25,   20.76){\circle*{1.0}}
\put(  29.50,   21.06){\circle*{1.0}}
\put(  29.75,   21.36){\circle*{1.0}}
\put(  30.00,   21.67){\circle*{1.0}}
\put(  30.25,   21.97){\circle*{1.0}}
\put(  30.50,   22.26){\circle*{1.0}}
\put(  30.75,   22.56){\circle*{1.0}}
\put(  31.00,   22.87){\circle*{1.0}}
\put(  31.25,   23.16){\circle*{1.0}}
\put(  31.50,   23.46){\circle*{1.0}}
\put(  31.75,   23.76){\circle*{1.0}}
\put(  32.00,   24.07){\circle*{1.0}}
\put(  32.25,   24.36){\circle*{1.0}}

\put(  32.50,   24.66){\circle*{1.0}} 
\put(  32.75,   24.98){\circle*{1.0}} 
\put(  33.00,   25.31){\circle*{1.0}} 
\put(  33.25,   25.63){\circle*{1.0}} 
\put(  33.50,   25.95){\circle*{1.0}} 
\put(  33.75,   26.20){\circle*{1.0}} 
\put(  34.00,   26.52){\circle*{1.0}} 
\put(  34.25,   26.85){\circle*{1.0}} 
\put(  34.50,   27.17){\circle*{1.0}} 
\put(  34.75,   27.41){\circle*{1.0}} 
\put(  35.00,   27.74){\circle*{1.0}} 
\put(  35.25,   28.06){\circle*{1.0}} 
\put(  35.50,   28.39){\circle*{1.0}} 
\put(  35.75,   28.63){\circle*{1.0}} 
\put(  36.00,   28.95){\circle*{1.0}} 
\put(  36.25,   29.28){\circle*{1.0}} 
\put(  36.50,   29.52){\circle*{1.0}} 
\put(  36.75,   29.84){\circle*{1.0}} 
\put(  37.00,   30.17){\circle*{1.0}} 
\put(  37.25,   30.49){\circle*{1.0}} 
\put(  37.50,   30.74){\circle*{1.0}} 
\put(  37.75,   31.06){\circle*{1.0}} 
\put(  38.00,   31.39){\circle*{1.0}} 
\put(  38.25,   31.63){\circle*{1.0}} 
\put(  38.50,   31.95){\circle*{1.0}} 
\put(  38.75,   32.28){\circle*{1.0}} 
\put(  39.00,   32.52){\circle*{1.0}} 
\put(  39.25,   32.84){\circle*{1.0}} 
\put(  39.50,   33.09){\circle*{1.0}} 
\put(  39.75,   33.41){\circle*{1.0}} 
\put(  40.00,   33.74){\circle*{1.0}} 
\put(  40.25,   33.98){\circle*{1.0}} 
\put(  40.50,   34.30){\circle*{1.0}} 
\put(  40.75,   34.55){\circle*{1.0}} 
\put(  41.00,   34.87){\circle*{1.0}} 
\put(  41.25,   35.11){\circle*{1.0}} 
\put(  41.50,   35.44){\circle*{1.0}} 
\put(  41.75,   35.68){\circle*{1.0}} 
\put(  42.00,   36.01){\circle*{1.0}} 
\put(  42.25,   36.25){\circle*{1.0}} 
\put(  42.50,   36.49){\circle*{1.0}} 
\put(  42.75,   36.82){\circle*{1.0}} 
\put(  43.00,   37.06){\circle*{1.0}} 
\put(  43.25,   37.30){\circle*{1.0}} 
\put(  43.50,   37.63){\circle*{1.0}} 
\put(  43.75,   37.87){\circle*{1.0}} 
\put(  44.00,   38.11){\circle*{1.0}} 
\put(  44.25,   38.44){\circle*{1.0}} 
\put(  44.50,   38.68){\circle*{1.0}} 
\put(  44.75,   38.92){\circle*{1.0}} 
\put(  45.00,   39.17){\circle*{1.0}} 
\put(  45.25,   39.41){\circle*{1.0}} 
\put(  45.50,   39.73){\circle*{1.0}} 
\put(  45.75,   39.98){\circle*{1.0}} 
\put(  46.00,   40.22){\circle*{1.0}} 
\put(  46.25,   40.46){\circle*{1.0}} 
\put(  46.50,   40.71){\circle*{1.0}} 
\put(  46.75,   40.95){\circle*{1.0}} 
\put(  47.00,   41.19){\circle*{1.0}} 
\put(  47.25,   41.44){\circle*{1.0}} 
\put(  47.50,   41.68){\circle*{1.0}} 
\put(  47.75,   41.92){\circle*{1.0}} 
\put(  48.00,   42.17){\circle*{1.0}} 
\put(  48.25,   42.33){\circle*{1.0}} 
\put(  48.50,   42.57){\circle*{1.0}} 
\put(  48.75,   42.81){\circle*{1.0}} 
\put(  49.00,   43.06){\circle*{1.0}} 
\put(  49.25,   43.30){\circle*{1.0}} 
\put(  49.50,   43.46){\circle*{1.0}} 
\put(  49.75,   43.71){\circle*{1.0}} 
\put(  50.00,   43.95){\circle*{1.0}} 
\put(  50.25,   44.11){\circle*{1.0}} 
\put(  50.50,   44.35){\circle*{1.0}} 
\put(  50.75,   44.52){\circle*{1.0}} 
\put(  51.00,   44.76){\circle*{1.0}} 
\put(  51.25,   45.00){\circle*{1.0}} 
\put(  51.50,   45.16){\circle*{1.0}} 
\put(  51.75,   45.33){\circle*{1.0}} 
\put(  52.00,   45.57){\circle*{1.0}} 
\put(  52.25,   45.73){\circle*{1.0}} 
\put(  52.50,   45.98){\circle*{1.0}} 
\put(  52.75,   46.14){\circle*{1.0}} 
\put(  53.00,   46.30){\circle*{1.0}} 
\put(  53.25,   46.46){\circle*{1.0}} 
\put(  53.50,   46.70){\circle*{1.0}} 
\put(  53.75,   46.87){\circle*{1.0}} 
\put(  54.00,   47.03){\circle*{1.0}} 
\put(  54.25,   47.19){\circle*{1.0}} 
\put(  54.50,   47.35){\circle*{1.0}} 
\put(  54.75,   47.52){\circle*{1.0}} 
\put(  55.00,   47.68){\circle*{1.0}} 
\put(  55.25,   47.84){\circle*{1.0}} 
\put(  55.50,   48.00){\circle*{1.0}} 
\put(  55.75,   48.16){\circle*{1.0}} 
\put(  56.00,   48.33){\circle*{1.0}} 
\put(  56.25,   48.49){\circle*{1.0}} 
\put(  56.50,   48.65){\circle*{1.0}} 
\put(  56.75,   48.81){\circle*{1.0}} 
\put(  57.00,   48.89){\circle*{1.0}} 
\put(  57.25,   49.06){\circle*{1.0}} 
\put(  57.50,   49.22){\circle*{1.0}} 
\put(  57.75,   49.38){\circle*{1.0}} 
\put(  58.00,   49.46){\circle*{1.0}} 
\put(  58.25,   49.62){\circle*{1.0}} 
\put(  58.50,   49.70){\circle*{1.0}} 
\put(  58.75,   49.87){\circle*{1.0}} 
\put(  59.00,   50.03){\circle*{1.0}} 
\put(  59.25,   50.11){\circle*{1.0}} 
\put(  59.50,   50.19){\circle*{1.0}} 
\put(  59.75,   50.35){\circle*{1.0}} 
\put(  60.00,   50.43){\circle*{1.0}} 
\put(  60.25,   50.60){\circle*{1.0}} 
\put(  60.50,   50.68){\circle*{1.0}} 
\put(  60.75,   50.76){\circle*{1.0}} 
\put(  61.00,   50.92){\circle*{1.0}} 
\put(  61.25,   51.00){\circle*{1.0}} 
\put(  61.50,   51.08){\circle*{1.0}} 
\put(  61.75,   51.16){\circle*{1.0}} 
\put(  62.00,   51.24){\circle*{1.0}} 
\put(  62.25,   51.41){\circle*{1.0}} 
\put(  62.50,   51.49){\circle*{1.0}} 
\put(  63.00,   51.57){\circle*{1.0}} 
\put(  63.25,   51.49){\circle*{1.0}} 
\put(  63.50,   51.41){\circle*{1.0}} 
\put(  63.75,   51.33){\circle*{1.0}} 
\put(  64.00,   51.24){\circle*{1.0}} 
\put(  64.25,   51.24){\circle*{1.0}} 
\put(  64.50,   51.16){\circle*{1.0}} 
\put(  64.75,   51.08){\circle*{1.0}} 
\put(  65.00,   51.08){\circle*{1.0}} 
\put(  65.25,   51.00){\circle*{1.0}} 
\put(  65.50,   50.92){\circle*{1.0}} 
\put(  65.75,   50.92){\circle*{1.0}} 
\put(  66.00,   50.84){\circle*{1.0}} 
\put(  66.25,   50.84){\circle*{1.0}} 
\put(  66.50,   50.76){\circle*{1.0}} 
\put(  66.75,   50.76){\circle*{1.0}} 
\put(  67.00,   50.68){\circle*{1.0}} 
\put(  67.25,   50.68){\circle*{1.0}} 
\put(  67.50,   50.60){\circle*{1.0}} 
\put(  67.75,   50.60){\circle*{1.0}} 
\put(  68.00,   50.51){\circle*{1.0}} 
\put(  68.25,   50.51){\circle*{1.0}} 
\put(  68.50,   50.51){\circle*{1.0}} 
\put(  68.75,   50.43){\circle*{1.0}} 
\put(  69.00,   50.43){\circle*{1.0}} 
\put(  69.25,   50.43){\circle*{1.0}} 
\put(  69.50,   50.35){\circle*{1.0}} 
\put(  69.75,   50.35){\circle*{1.0}} 
\put(  70.00,   50.35){\circle*{1.0}} 
\put(  70.25,   50.35){\circle*{1.0}} 
\put(  70.50,   50.35){\circle*{1.0}} 
\put(  70.75,   50.27){\circle*{1.0}} 
\put(  71.00,   50.27){\circle*{1.0}} 
\put(  71.25,   50.27){\circle*{1.0}} 
\put(  71.50,   50.27){\circle*{1.0}} 
\put(  71.75,   50.27){\circle*{1.0}} 
\put(  72.00,   50.27){\circle*{1.0}} 
\put(  72.25,   50.27){\circle*{1.0}} 
\put(  72.50,   50.27){\circle*{1.0}} 
\put(  72.75,   50.27){\circle*{1.0}} 
\put(  73.00,   50.27){\circle*{1.0}} 
\put(  73.25,   50.19){\circle*{1.0}} 
\put(  73.50,   50.19){\circle*{1.0}} 
\put(  73.75,   50.27){\circle*{1.0}} 
\put(  74.00,   50.27){\circle*{1.0}} 
\put(  74.25,   50.27){\circle*{1.0}} 
\put(  74.50,   50.27){\circle*{1.0}} 
\put(  74.75,   50.27){\circle*{1.0}} 
\put(  75.00,   50.27){\circle*{1.0}} 
\put(  75.25,   50.27){\circle*{1.0}} 
\put(  75.50,   50.27){\circle*{1.0}} 
\put(  75.75,   50.27){\circle*{1.0}} 
\put(  76.00,   50.27){\circle*{1.0}} 
\put(  76.25,   50.27){\circle*{1.0}} 
\put(  76.50,   50.27){\circle*{1.0}} 
\put(  76.75,   50.35){\circle*{1.0}} 
\put(  77.00,   50.35){\circle*{1.0}} 
\put(  77.25,   50.35){\circle*{1.0}} 
\put(  77.50,   50.35){\circle*{1.0}} 
\put(  77.75,   50.35){\circle*{1.0}} 
\put(  78.00,   50.43){\circle*{1.0}} 
\put(  78.25,   50.43){\circle*{1.0}} 
\put(  78.50,   50.43){\circle*{1.0}} 
\put(  78.75,   50.43){\circle*{1.0}} 
\put(  79.00,   50.43){\circle*{1.0}} 
\put(  79.25,   50.51){\circle*{1.0}} 
\put(  79.50,   50.51){\circle*{1.0}} 
\put(  79.75,   50.51){\circle*{1.0}} 
\put(  80.00,   50.51){\circle*{1.0}} 
\put(  80.25,   50.60){\circle*{1.0}} 
\put(  80.50,   50.60){\circle*{1.0}} 
\put(  80.75,   50.60){\circle*{1.0}} 
\put(  81.00,   50.60){\circle*{1.0}} 
\put(  81.25,   50.68){\circle*{1.0}} 
\put(  81.50,   50.68){\circle*{1.0}} 
\put(  81.75,   50.68){\circle*{1.0}} 
\put(  82.00,   50.76){\circle*{1.0}} 
\put(  82.25,   50.76){\circle*{1.0}} 
\put(  82.50,   50.76){\circle*{1.0}} 
\put(  82.75,   50.84){\circle*{1.0}} 
\put(  83.00,   50.84){\circle*{1.0}} 
\put(  83.25,   50.84){\circle*{1.0}} 
\put(  83.50,   50.92){\circle*{1.0}} 
\put(  83.75,   50.92){\circle*{1.0}} 
\put(  84.00,   50.92){\circle*{1.0}} 
\put(  84.25,   50.92){\circle*{1.0}} 
\put(  84.50,   51.00){\circle*{1.0}} 
\put(  84.75,   51.00){\circle*{1.0}} 
\put(  85.00,   51.00){\circle*{1.0}} 
\put(  85.25,   51.08){\circle*{1.0}} 
\put(  85.50,   51.08){\circle*{1.0}} 
\put(  85.75,   51.08){\circle*{1.0}} 
\put(  86.00,   51.16){\circle*{1.0}} 
\put(  86.25,   51.16){\circle*{1.0}} 
\put(  86.50,   51.16){\circle*{1.0}} 
\put(  86.75,   51.24){\circle*{1.0}} 
\put(  87.00,   51.24){\circle*{1.0}} 
\put(  87.25,   51.24){\circle*{1.0}} 
\put(  87.50,   51.33){\circle*{1.0}} 
\put(  87.75,   51.33){\circle*{1.0}} 
\put(  88.00,   51.33){\circle*{1.0}} 
\put(  88.25,   51.41){\circle*{1.0}} 
\put(  88.50,   51.41){\circle*{1.0}} 
\put(  88.75,   51.41){\circle*{1.0}} 
\put(  89.00,   51.49){\circle*{1.0}} 
\put(  89.25,   51.49){\circle*{1.0}} 
\put(  89.50,   51.49){\circle*{1.0}} 
\put(  89.75,   51.57){\circle*{1.0}} 
\put(  90.00,   51.57){\circle*{1.0}} 
\put(  90.25,   51.57){\circle*{1.0}} 
\put(  90.50,   51.65){\circle*{1.0}} 
\put(  90.75,   51.65){\circle*{1.0}} 
\put(  91.00,   51.65){\circle*{1.0}} 
\put(  91.25,   51.73){\circle*{1.0}} 
\put(  91.50,   51.73){\circle*{1.0}} 
\put(  91.75,   51.73){\circle*{1.0}} 
\put(  92.00,   51.81){\circle*{1.0}} 
\put(  92.25,   51.81){\circle*{1.0}} 
\put(  92.50,   51.81){\circle*{1.0}} 
\put(  92.75,   51.81){\circle*{1.0}} 
\put(  93.00,   51.89){\circle*{1.0}} 
\put(  93.25,   51.89){\circle*{1.0}} 
\put(  93.50,   51.89){\circle*{1.0}} 
\put(  93.75,   51.89){\circle*{1.0}}
\put(  94.00,   51.89){\circle*{1.0}}
\put(  94.25,   51.89){\circle*{1.0}}
\put(  94.50,   51.89){\circle*{1.0}}
\put(  94.75,   51.89){\circle*{1.0}}
\put(  95.00,   51.89){\circle*{1.0}}
\put(  95.25,   51.89){\circle*{1.0}} 
\put(  95.50,   51.89){\circle*{1.0}} 
\put(  95.75,   51.89){\circle*{1.0}} 
\put(  96.00,   51.89){\circle*{1.0}} 
\put(  96.25,   51.81){\circle*{1.0}} 
\put(  96.50,   51.81){\circle*{1.0}} 
\put(  96.75,   51.81){\circle*{1.0}} 
\put(  97.00,   51.81){\circle*{1.0}} 
\put(  97.25,   51.81){\circle*{1.0}} 
\put(  97.50,   51.73){\circle*{1.0}} 
\put(  97.75,   51.73){\circle*{1.0}} 
\put(  98.00,   51.73){\circle*{1.0}} 
\put(  98.25,   51.73){\circle*{1.0}} 
\put(  98.50,   51.73){\circle*{1.0}} 
\put(  98.75,   51.73){\circle*{1.0}} 
\put(  99.00,   51.73){\circle*{1.0}} 
\put(  99.25,   51.65){\circle*{1.0}} 
\put(  99.50,   51.65){\circle*{1.0}} 
\put(  99.75,   51.65){\circle*{1.0}} 
\put( 100.00,   51.65){\circle*{1.0}} 

\put(   0.25,    0.01){\circle*{1.0}} 
\put(   0.50,    0.05){\circle*{1.0}} 
\put(   0.75,    0.03){\circle*{1.0}} 
\put(   1.00,    0.06){\circle*{1.0}} 
\put(   1.25,    0.07){\circle*{1.0}} 
\put(   1.50,    0.13){\circle*{1.0}} 
\put(   1.75,    0.14){\circle*{1.0}} 
\put(   2.00,    0.18){\circle*{1.0}} 
\put(   2.25,    0.22){\circle*{1.0}} 
\put(   2.50,    0.25){\circle*{1.0}} 
\put(   2.75,    0.32){\circle*{1.0}} 
\put(   3.00,    0.38){\circle*{1.0}} 
\put(   3.25,    0.44){\circle*{1.0}} 
\put(   3.50,    0.49){\circle*{1.0}} 
\put(   3.75,    0.58){\circle*{1.0}} 
\put(   4.00,    0.66){\circle*{1.0}} 
\put(   4.25,    0.74){\circle*{1.0}} 
\put(   4.50,    0.82){\circle*{1.0}} 
\put(   4.75,    0.88){\circle*{1.0}} 
\put(   5.00,    0.99){\circle*{1.0}} 
\put(   5.25,    1.09){\circle*{1.0}} 
\put(   5.50,    1.22){\circle*{1.0}} 
\put(   5.75,    1.31){\circle*{1.0}} 
\put(   6.00,    1.43){\circle*{1.0}} 
\put(   6.25,    1.55){\circle*{1.0}} 
\put(   6.50,    1.67){\circle*{1.0}} 
\put(   6.75,    1.82){\circle*{1.0}} 
\put(   7.00,    1.92){\circle*{1.0}} 
\put(   7.25,    2.06){\circle*{1.0}} 
\put(   7.50,    2.19){\circle*{1.0}} 
\put(   7.75,    2.36){\circle*{1.0}} 
\put(   8.00,    2.53){\circle*{1.0}} 
\put(   8.25,    2.69){\circle*{1.0}} 
\put(   8.50,    2.84){\circle*{1.0}} 
\put(   8.75,    2.99){\circle*{1.0}} 
\put(   9.00,    3.18){\circle*{1.0}} 
\put(   9.25,    3.36){\circle*{1.0}} 
\put(   9.50,    3.53){\circle*{1.0}} 
\put(   9.75,    3.70){\circle*{1.0}} 
\put(  10.00,    3.91){\circle*{1.0}} 
\put(  10.25,    4.11){\circle*{1.0}} 
\put(  10.50,    4.30){\circle*{1.0}} 
\put(  10.75,    4.49){\circle*{1.0}} 
\put(  11.00,    4.72){\circle*{1.0}} 
\put(  11.25,    4.90){\circle*{1.0}} 
\put(  11.50,    5.11){\circle*{1.0}} 
\put(  11.75,    5.36){\circle*{1.0}} 
\put(  12.00,    5.57){\circle*{1.0}} 
\put(  12.25,    5.81){\circle*{1.0}} 
\put(  12.50,    6.04){\circle*{1.0}} 
\put(  12.75,    6.28){\circle*{1.0}} 
\put(  13.00,    6.50){\circle*{1.0}} 
\put(  13.25,    6.76){\circle*{1.0}} 
\put(  13.50,    7.02){\circle*{1.0}} 
\put(  13.75,    7.27){\circle*{1.0}} 
\put(  14.00,    7.51){\circle*{1.0}} 
\put(  14.25,    7.79){\circle*{1.0}} 
\put(  14.50,    8.07){\circle*{1.0}} 
\put(  14.75,    8.34){\circle*{1.0}} 
\put(  15.00,    8.61){\circle*{1.0}} 
\put(  15.25,    8.87){\circle*{1.0}} 
\put(  15.50,    9.17){\circle*{1.0}} 
\put(  15.75,    9.46){\circle*{1.0}} 
\put(  16.00,    9.74){\circle*{1.0}} 
\put(  16.25,   10.02){\circle*{1.0}} 
\put(  16.50,   10.34){\circle*{1.0}} 
\put(  16.75,   10.65){\circle*{1.0}} 
\put(  17.00,   10.96){\circle*{1.0}} 
\put(  17.25,   11.26){\circle*{1.0}} 
\put(  17.50,   11.60){\circle*{1.0}} 
\put(  17.75,   11.89){\circle*{1.0}} 
\put(  18.00,   12.22){\circle*{1.0}} 
\put(  18.25,   12.54){\circle*{1.0}} 
\put(  18.50,   12.89){\circle*{1.0}} 
\put(  18.75,   13.21){\circle*{1.0}} 
\put(  19.00,   13.55){\circle*{1.0}} 
\put(  19.25,   13.89){\circle*{1.0}} 
\put(  19.50,   14.23){\circle*{1.0}} 
\put(  19.75,   14.56){\circle*{1.0}} 
\put(  20.00,   14.93){\circle*{1.0}} 
\put(  20.25,   15.29){\circle*{1.0}} 
\put(  20.50,   15.65){\circle*{1.0}} 
\put(  20.75,   16.00){\circle*{1.0}} 
\put(  21.00,   16.39){\circle*{1.0}} 
\put(  21.25,   16.73){\circle*{1.0}} 
\put(  21.50,   17.11){\circle*{1.0}} 
\put(  21.75,   17.48){\circle*{1.0}} 
\put(  22.00,   17.85){\circle*{1.0}} 
\put(  22.25,   18.25){\circle*{1.0}} 
\put(  22.50,   18.61){\circle*{1.0}} 
\put(  22.75,   19.00){\circle*{1.0}} 
\put(  23.00,   19.39){\circle*{1.0}} 
\put(  23.25,   19.77){\circle*{1.0}} 
\put(  23.50,   20.19){\circle*{1.0}} 
\put(  23.75,   20.60){\circle*{1.0}} 
\put(  24.00,   20.97){\circle*{1.0}} 
\put(  24.25,   21.37){\circle*{1.0}} 
\put(  24.50,   21.77){\circle*{1.0}} 
\put(  24.75,   22.20){\circle*{1.0}} 
\put(  25.00,   22.59){\circle*{1.0}} 
\put(  25.25,   23.01){\circle*{1.0}} 
\put(  25.50,   23.43){\circle*{1.0}} 
\put(  25.75,   23.84){\circle*{1.0}} 
\put(  26.00,   24.29){\circle*{1.0}} 

\put(  26.25,   24.73){\circle*{1.0}}
\put(  26.50,   25.18){\circle*{1.0}}
\put(  26.75,   25.62){\circle*{1.0}}
\put(  27.00,   26.07){\circle*{1.0}}
\put(  27.25,   26.51){\circle*{1.0}}
\put(  27.50,   26.96){\circle*{1.0}}
\put(  27.75,   27.40){\circle*{1.0}}
\put(  28.00,   27.89){\circle*{1.0}}
\put(  28.25,   28.29){\circle*{1.0}}
\put(  28.50,   28.73){\circle*{1.0}}

\put(  28.75,   29.18){\circle*{1.0}}
\put(  29.00,   29.62){\circle*{1.0}}
\put(  29.25,   30.07){\circle*{1.0}}
\put(  29.50,   30.51){\circle*{1.0}}
\put(  29.75,   30.95){\circle*{1.0}}
\put(  30.00,   31.40){\circle*{1.0}}
\put(  30.25,   31.84){\circle*{1.0}}
\put(  30.50,   32.29){\circle*{1.0}}
\put(  30.75,   32.73){\circle*{1.0}}
\put(  31.00,   33.17){\circle*{1.0}}
\put(  31.25,   33.62){\circle*{1.0}}
\put(  31.50,   34.06){\circle*{1.0}}
\put(  31.75,   34.51){\circle*{1.0}}
\put(  32.00,   34.95){\circle*{1.0}}
\put(  32.25,   35.40){\circle*{1.0}}
\put(  32.50,   35.84){\circle*{1.0}}
\put(  32.75,   36.29){\circle*{1.0}}

\put(  33.00,   36.73){\circle*{1.0}}
\put(  33.25,   37.22){\circle*{1.0}} 
\put(  33.50,   37.71){\circle*{1.0}} 
\put(  33.75,   38.11){\circle*{1.0}} 
\put(  34.00,   38.60){\circle*{1.0}} 
\put(  34.25,   39.09){\circle*{1.0}} 
\put(  34.50,   39.49){\circle*{1.0}} 
\put(  34.75,   39.98){\circle*{1.0}} 
\put(  35.00,   40.46){\circle*{1.0}} 
\put(  35.25,   40.95){\circle*{1.0}} 
\put(  35.50,   41.36){\circle*{1.0}} 
\put(  35.75,   41.84){\circle*{1.0}} 
\put(  36.00,   42.33){\circle*{1.0}} 
\put(  36.25,   42.73){\circle*{1.0}} 
\put(  36.50,   43.22){\circle*{1.0}} 
\put(  36.75,   43.71){\circle*{1.0}} 
\put(  37.00,   44.11){\circle*{1.0}} 
\put(  37.25,   44.60){\circle*{1.0}} 
\put(  37.50,   45.08){\circle*{1.0}} 
\put(  37.75,   45.49){\circle*{1.0}} 
\put(  38.00,   45.98){\circle*{1.0}} 
\put(  38.25,   46.46){\circle*{1.0}} 
\put(  38.50,   46.87){\circle*{1.0}} 
\put(  38.75,   47.35){\circle*{1.0}} 
\put(  39.00,   47.84){\circle*{1.0}} 
\put(  39.25,   48.24){\circle*{1.0}} 
\put(  39.50,   48.73){\circle*{1.0}} 
\put(  39.75,   49.14){\circle*{1.0}} 
\put(  40.00,   49.62){\circle*{1.0}} 
\put(  40.25,   50.03){\circle*{1.0}} 
\put(  40.50,   50.51){\circle*{1.0}} 
\put(  40.75,   50.92){\circle*{1.0}} 
\put(  41.00,   51.41){\circle*{1.0}} 
\put(  41.25,   51.81){\circle*{1.0}} 
\put(  41.50,   52.30){\circle*{1.0}} 
\put(  41.75,   52.70){\circle*{1.0}} 
\put(  42.00,   53.19){\circle*{1.0}} 
\put(  42.25,   53.59){\circle*{1.0}} 
\put(  42.50,   54.00){\circle*{1.0}} 
\put(  42.75,   54.49){\circle*{1.0}} 
\put(  43.00,   54.89){\circle*{1.0}} 
\put(  43.25,   55.30){\circle*{1.0}} 
\put(  43.50,   55.70){\circle*{1.0}} 
\put(  43.75,   56.19){\circle*{1.0}} 
\put(  44.00,   56.59){\circle*{1.0}} 
\put(  44.25,   57.00){\circle*{1.0}} 
\put(  44.50,   57.40){\circle*{1.0}} 
\put(  44.75,   57.81){\circle*{1.0}} 
\put(  45.00,   58.21){\circle*{1.0}} 
\put(  45.25,   58.62){\circle*{1.0}} 
\put(  45.50,   59.03){\circle*{1.0}} 
\put(  45.75,   59.43){\circle*{1.0}} 
\put(  46.00,   59.84){\circle*{1.0}} 
\put(  46.25,   60.24){\circle*{1.0}} 
\put(  46.50,   60.65){\circle*{1.0}} 
\put(  46.75,   61.05){\circle*{1.0}} 
\put(  47.00,   61.46){\circle*{1.0}} 
\put(  47.25,   61.78){\circle*{1.0}} 
\put(  47.50,   62.19){\circle*{1.0}} 
\put(  47.75,   62.59){\circle*{1.0}} 
\put(  48.00,   62.92){\circle*{1.0}} 
\put(  48.25,   63.32){\circle*{1.0}} 
\put(  48.50,   63.65){\circle*{1.0}} 
\put(  48.75,   64.05){\circle*{1.0}} 
\put(  49.00,   64.38){\circle*{1.0}} 
\put(  49.25,   64.78){\circle*{1.0}} 
\put(  49.50,   65.10){\circle*{1.0}} 
\put(  49.75,   65.51){\circle*{1.0}} 
\put(  50.00,   65.83){\circle*{1.0}} 
\put(  50.25,   66.16){\circle*{1.0}} 
\put(  50.50,   66.48){\circle*{1.0}} 
\put(  50.75,   66.89){\circle*{1.0}} 
\put(  51.00,   67.21){\circle*{1.0}} 
\put(  51.25,   67.54){\circle*{1.0}} 
\put(  51.50,   67.86){\circle*{1.0}} 
\put(  51.75,   68.18){\circle*{1.0}} 
\put(  52.00,   68.51){\circle*{1.0}} 
\put(  52.25,   68.83){\circle*{1.0}} 
\put(  52.50,   69.08){\circle*{1.0}} 
\put(  52.75,   69.40){\circle*{1.0}} 
\put(  53.00,   69.73){\circle*{1.0}} 
\put(  53.25,   70.05){\circle*{1.0}} 
\put(  53.50,   70.29){\circle*{1.0}} 
\put(  53.75,   70.62){\circle*{1.0}} 
\put(  54.00,   70.86){\circle*{1.0}} 
\put(  54.25,   71.18){\circle*{1.0}} 
\put(  54.50,   71.43){\circle*{1.0}} 
\put(  54.75,   71.75){\circle*{1.0}} 
\put(  55.00,   71.99){\circle*{1.0}} 
\put(  55.25,   72.24){\circle*{1.0}} 
\put(  55.50,   72.56){\circle*{1.0}} 
\put(  55.75,   72.81){\circle*{1.0}} 
\put(  56.00,   73.05){\circle*{1.0}} 
\put(  56.25,   73.29){\circle*{1.0}} 
\put(  56.50,   73.53){\circle*{1.0}} 
\put(  56.75,   73.78){\circle*{1.0}} 
\put(  57.00,   74.02){\circle*{1.0}} 
\put(  57.25,   74.26){\circle*{1.0}} 
\put(  57.50,   74.51){\circle*{1.0}} 
\put(  57.75,   74.75){\circle*{1.0}} 
\put(  58.00,   74.91){\circle*{1.0}} 
\put(  58.25,   75.16){\circle*{1.0}} 
\put(  58.50,   75.40){\circle*{1.0}} 
\put(  58.75,   75.56){\circle*{1.0}} 
\put(  59.00,   75.80){\circle*{1.0}} 
\put(  59.25,   75.97){\circle*{1.0}} 
\put(  59.50,   76.21){\circle*{1.0}} 
\put(  59.75,   76.37){\circle*{1.0}} 
\put(  60.00,   76.53){\circle*{1.0}} 
\put(  60.25,   76.78){\circle*{1.0}} 
\put(  60.50,   76.94){\circle*{1.0}} 
\put(  60.75,   77.10){\circle*{1.0}} 
\put(  61.00,   77.26){\circle*{1.0}} 
\put(  61.25,   77.43){\circle*{1.0}} 
\put(  61.50,   77.59){\circle*{1.0}} 
\put(  61.75,   77.75){\circle*{1.0}} 
\put(  62.00,   77.91){\circle*{1.0}} 
\put(  62.25,   78.07){\circle*{1.0}} 
\put(  62.50,   78.24){\circle*{1.0}} 
\put(  62.75,   78.40){\circle*{1.0}} 
\put(  63.00,   78.40){\circle*{1.0}} 
\put(  63.25,   78.32){\circle*{1.0}} 
\put(  63.50,   78.15){\circle*{1.0}} 
\put(  63.75,   78.07){\circle*{1.0}} 
\put(  64.00,   77.99){\circle*{1.0}} 
\put(  64.25,   77.91){\circle*{1.0}} 
\put(  64.50,   77.83){\circle*{1.0}} 
\put(  64.75,   77.75){\circle*{1.0}} 
\put(  65.00,   77.67){\circle*{1.0}} 
\put(  65.25,   77.59){\circle*{1.0}} 
\put(  65.50,   77.51){\circle*{1.0}} 
\put(  65.75,   77.43){\circle*{1.0}} 
\put(  66.00,   77.34){\circle*{1.0}} 
\put(  66.25,   77.26){\circle*{1.0}} 
\put(  66.50,   77.18){\circle*{1.0}} 
\put(  66.75,   77.18){\circle*{1.0}} 
\put(  67.00,   77.10){\circle*{1.0}} 
\put(  67.25,   77.02){\circle*{1.0}} 
\put(  67.50,   77.02){\circle*{1.0}} 
\put(  67.75,   76.94){\circle*{1.0}} 
\put(  68.00,   76.86){\circle*{1.0}} 
\put(  68.25,   76.86){\circle*{1.0}} 
\put(  68.50,   76.78){\circle*{1.0}} 
\put(  68.75,   76.78){\circle*{1.0}} 
\put(  69.00,   76.70){\circle*{1.0}} 
\put(  69.25,   76.70){\circle*{1.0}} 
\put(  69.50,   76.70){\circle*{1.0}} 
\put(  69.75,   76.61){\circle*{1.0}} 
\put(  70.00,   76.61){\circle*{1.0}} 
\put(  70.25,   76.61){\circle*{1.0}} 
\put(  70.50,   76.53){\circle*{1.0}} 
\put(  70.75,   76.53){\circle*{1.0}} 
\put(  71.00,   76.53){\circle*{1.0}} 
\put(  71.25,   76.53){\circle*{1.0}} 
\put(  71.50,   76.53){\circle*{1.0}} 
\put(  71.75,   76.45){\circle*{1.0}} 
\put(  72.00,   76.45){\circle*{1.0}} 
\put(  72.25,   76.45){\circle*{1.0}} 
\put(  72.50,   76.45){\circle*{1.0}} 
\put(  72.75,   76.45){\circle*{1.0}} 
\put(  73.00,   76.45){\circle*{1.0}} 
\put(  73.25,   76.45){\circle*{1.0}} 
\put(  73.50,   76.45){\circle*{1.0}} 
\put(  73.75,   76.45){\circle*{1.0}} 
\put(  74.00,   76.45){\circle*{1.0}} 
\put(  74.25,   76.45){\circle*{1.0}} 
\put(  74.50,   76.45){\circle*{1.0}} 
\put(  74.75,   76.53){\circle*{1.0}} 
\put(  75.00,   76.53){\circle*{1.0}} 
\put(  75.25,   76.53){\circle*{1.0}} 
\put(  75.50,   76.53){\circle*{1.0}} 
\put(  75.75,   76.53){\circle*{1.0}} 
\put(  76.00,   76.61){\circle*{1.0}} 
\put(  76.25,   76.61){\circle*{1.0}} 
\put(  76.50,   76.61){\circle*{1.0}} 
\put(  76.75,   76.61){\circle*{1.0}} 
\put(  77.00,   76.70){\circle*{1.0}} 
\put(  77.25,   76.70){\circle*{1.0}} 
\put(  77.50,   76.70){\circle*{1.0}} 
\put(  77.75,   76.78){\circle*{1.0}} 
\put(  78.00,   76.78){\circle*{1.0}} 
\put(  78.25,   76.78){\circle*{1.0}} 
\put(  78.50,   76.86){\circle*{1.0}} 
\put(  78.75,   76.86){\circle*{1.0}} 
\put(  79.00,   76.86){\circle*{1.0}} 
\put(  79.25,   76.94){\circle*{1.0}} 
\put(  79.50,   76.94){\circle*{1.0}} 
\put(  79.75,   77.02){\circle*{1.0}} 
\put(  80.00,   77.02){\circle*{1.0}} 
\put(  80.25,   77.10){\circle*{1.0}} 
\put(  80.50,   77.10){\circle*{1.0}} 
\put(  80.75,   77.18){\circle*{1.0}} 
\put(  81.00,   77.18){\circle*{1.0}} 
\put(  81.25,   77.26){\circle*{1.0}} 
\put(  81.50,   77.26){\circle*{1.0}} 
\put(  81.75,   77.34){\circle*{1.0}} 
\put(  82.00,   77.34){\circle*{1.0}} 
\put(  82.25,   77.43){\circle*{1.0}} 
\put(  82.50,   77.43){\circle*{1.0}} 
\put(  82.75,   77.51){\circle*{1.0}} 
\put(  83.00,   77.51){\circle*{1.0}} 
\put(  83.25,   77.59){\circle*{1.0}} 
\put(  83.50,   77.59){\circle*{1.0}} 
\put(  83.75,   77.67){\circle*{1.0}} 
\put(  84.00,   77.67){\circle*{1.0}} 
\put(  84.25,   77.75){\circle*{1.0}} 
\put(  84.50,   77.75){\circle*{1.0}} 
\put(  84.75,   77.83){\circle*{1.0}} 
\put(  85.00,   77.83){\circle*{1.0}} 
\put(  85.25,   77.91){\circle*{1.0}} 
\put(  85.50,   77.99){\circle*{1.0}} 
\put(  85.75,   77.99){\circle*{1.0}} 
\put(  86.00,   78.07){\circle*{1.0}} 
\put(  86.25,   78.07){\circle*{1.0}} 
\put(  86.50,   78.15){\circle*{1.0}} 
\put(  86.75,   78.15){\circle*{1.0}} 
\put(  87.00,   78.24){\circle*{1.0}} 
\put(  87.25,   78.32){\circle*{1.0}} 
\put(  87.50,   78.32){\circle*{1.0}} 
\put(  87.75,   78.40){\circle*{1.0}} 
\put(  88.00,   78.40){\circle*{1.0}} 
\put(  88.25,   78.48){\circle*{1.0}} 
\put(  88.50,   78.48){\circle*{1.0}} 
\put(  88.75,   78.56){\circle*{1.0}} 
\put(  89.00,   78.56){\circle*{1.0}} 
\put(  89.25,   78.64){\circle*{1.0}} 
\put(  89.50,   78.72){\circle*{1.0}} 
\put(  89.75,   78.72){\circle*{1.0}} 
\put(  90.00,   78.80){\circle*{1.0}} 
\put(  90.25,   78.80){\circle*{1.0}} 
\put(  90.50,   78.88){\circle*{1.0}} 
\put(  90.75,   78.88){\circle*{1.0}} 
\put(  91.00,   78.97){\circle*{1.0}} 
\put(  91.25,   78.97){\circle*{1.0}} 
\put(  91.50,   79.05){\circle*{1.0}} 
\put(  91.75,   79.05){\circle*{1.0}} 
\put(  92.00,   79.13){\circle*{1.0}} 
\put(  92.25,   79.13){\circle*{1.0}} 
\put(  92.50,   79.21){\circle*{1.0}} 
\put(  92.75,   79.21){\circle*{1.0}} 
\put(  93.00,   79.29){\circle*{1.0}} 
\put(  93.25,   79.29){\circle*{1.0}} 
\put(  93.50,   79.37){\circle*{1.0}} 
\put(  93.75,   79.37){\circle*{1.0}} 
\put(  94.00,   79.37){\circle*{1.0}}
\put(  94.25,   79.29){\circle*{1.0}}
\put(  94.50,   79.37){\circle*{1.0}}
\put(  94.75,   79.37){\circle*{1.0}}
\put(  95.00,   79.37){\circle*{1.0}}
\put(  95.25,   79.37){\circle*{1.0}}
\put(  95.50,   79.29){\circle*{1.0}}
\put(  95.75,   79.29){\circle*{1.0}} 
\put(  96.00,   79.29){\circle*{1.0}} 
\put(  96.25,   79.21){\circle*{1.0}} 
\put(  96.50,   79.21){\circle*{1.0}} 
\put(  96.75,   79.21){\circle*{1.0}} 
\put(  97.00,   79.13){\circle*{1.0}} 
\put(  97.25,   79.13){\circle*{1.0}} 
\put(  97.50,   79.13){\circle*{1.0}} 
\put(  97.75,   79.13){\circle*{1.0}} 
\put(  98.00,   79.05){\circle*{1.0}} 
\put(  98.25,   79.05){\circle*{1.0}} 
\put(  98.50,   79.05){\circle*{1.0}} 
\put(  98.75,   79.05){\circle*{1.0}} 
\put(  99.00,   79.05){\circle*{1.0}} 
\put(  99.25,   78.97){\circle*{1.0}} 
\put(  99.50,   78.97){\circle*{1.0}} 
\put(  99.75,   78.97){\circle*{1.0}} 
\put( 100.00,   78.97){\circle*{1.0}} 

\end{picture}
\end{center}
\vspace*{10 mm}

\caption{The critical load $p^{\ast}$ versus the total twist angle
$\Omega(L)$.
Curve~1: $\nu=1.1$;
curve~2: $\nu=1.3$;
curve~3: $\nu=1.5$}
\end{figure}

\end{document}